\documentclass[sigconf]{acmart}

\AtBeginDocument{%
  }

\copyrightyear{2024} 
\acmYear{2024} 
\setcopyright{rightsretained} 
\acmConference[ICMI Companion '24]{INTERNATIONAL CONFERENCE ON
MULTIMODAL INTERACTION}{November 4--8, 2024}{San Jose, Costa Rica}
\acmBooktitle{INTERNATIONAL CONFERENCE ON MULTIMODAL INTERACTION
(ICMI Companion '24), November 4--8, 2024, San Jose, Costa
Rica}\acmDOI{10.1145/3686215.3690152}
\acmISBN{979-8-4007-0463-5/24/11}
\begin{document}

\title{``Is This It?'': Towards Ecologically Valid Benchmarks \\
for Situated Collaboration}


\author{Dan Bohus}
\affiliation{%
  \institution{Microsoft Research}
  \city{Redmond}
  \country{United States}}
\email{dbohus@microsoft.com}
\orcid{0000-0002-6283-0590}

\author{Sean Andrist}
\affiliation{%
  \institution{Microsoft Research}
  \city{Redmond}
  \country{United States}}
\email{sandrist@microsoft.com}
\orcid{0000-0003-4972-7027}

\author{Yuwei Bao}
\authornote{Work conducted during an internship at Microsoft Research.}
\affiliation{%
  \institution{University of Michigan}
  \city{Ann Arbor}
  \country{United States}}
\email{yuweibao@umich.edu}
\orcid{0000-0002-7500-5944}

\author{Eric Horvitz}
\affiliation{%
  \institution{Microsoft}
  \city{Redmond}
  \country{United States}}
\email{horvitz@microsoft.com}
\orcid{0000-0002-8823-0614}

\author{Ann Paradiso}
\affiliation{%
  \institution{Microsoft Research}
  \city{Redmond}
  \country{United States}}
\email{annpar@microsoft.com}
\orcid{0000-0002-5092-8603}

% \renewcommand{\shortauthors}{Bohus et al.}
\renewcommand{\shorttitle}{``Is This It?'': Towards Ecologically Valid Benchmarks for Situated Collaboration}

\begin{abstract}
We report initial work towards constructing ecologically valid benchmarks to assess the capabilities of large multimodal models for engaging in situated collaboration. In contrast to existing benchmarks, in which question-answer pairs are generated post hoc over preexisting or synthetic datasets via templates, human annotators, or large language models (LLMs), we propose and investigate an interactive system-driven approach, where the questions are generated by users in context, during their interactions with an end-to-end situated AI system. We illustrate how the questions that arise are different in form and content from questions typically found in existing embodied question answering (EQA) benchmarks and discuss new real-world challenge problems brought to the fore.
\end{abstract}



\keywords{Embodied Question Answering; Benchmark Datasets; Mixed-Reality Task Assistance; Situated Collaboration}


\maketitle

\section{Introduction}

By coupling the generalist abilities of large language models with the capacity to ``see,'' large multimodal models (LMMs) \cite{gpt4v,phi,llava,kosmos2} promise to enable a wide array of AI-powered assistance scenarios in the physical world. Numerous applications can be envisioned, including collaborative robots, intelligent monitoring of factory floors, and mixed-reality glasses providing assistance on the fly.

There are still significant gaps between demonstrations of capabilities and the requirements for practical deployment in real-world scenarios. To track the performance of emerging models and understand their capabilities, the research community has developed a variety of benchmarks for video- and embodied- question answering \cite{openeqa,sq3d,leo,mmbench,seed,sok,star}. These benchmarks are typically constructed by identifying a preexisting multimodal dataset (or creating a synthetic one via a virtual environment), and then generating question-answer pairs from templates, human annotators, or via LLMs. The questions are designed to probe model capabilities along various dimensions, such as spatial understanding, episodic memory, and the recognition of objects and their attributes. While these benchmarks provide useful probes for model competency, we argue that they do not accurately capture the types of questions users ask when engaged in a real-time task, and thus do not reflect the full range of challenges that arise during situated collaboration.

To address this issue, we investigate an \emph{interactive system-driven} approach for constructing benchmarks and assessing capabilities of large multimodal models in situated assistance tasks. Rather than starting with a preexisting dataset and creating questions post hoc, we collect data with an end-to-end interactive AI system, and use it to identify challenges and define benchmarks. As we discuss below, this approach leads to different types of questions, and helps highlight novel embodied interaction challenges that go well beyond question answering. We argue that this approach provides a more ecologically valid and useful assessment of LMM capabilities, and that it can serve as a vehicle for real-world, continuous testing of models against new data.

\section{Related Work}

\begin{figure*}[t]
  \centering
  \includegraphics[width=\linewidth]{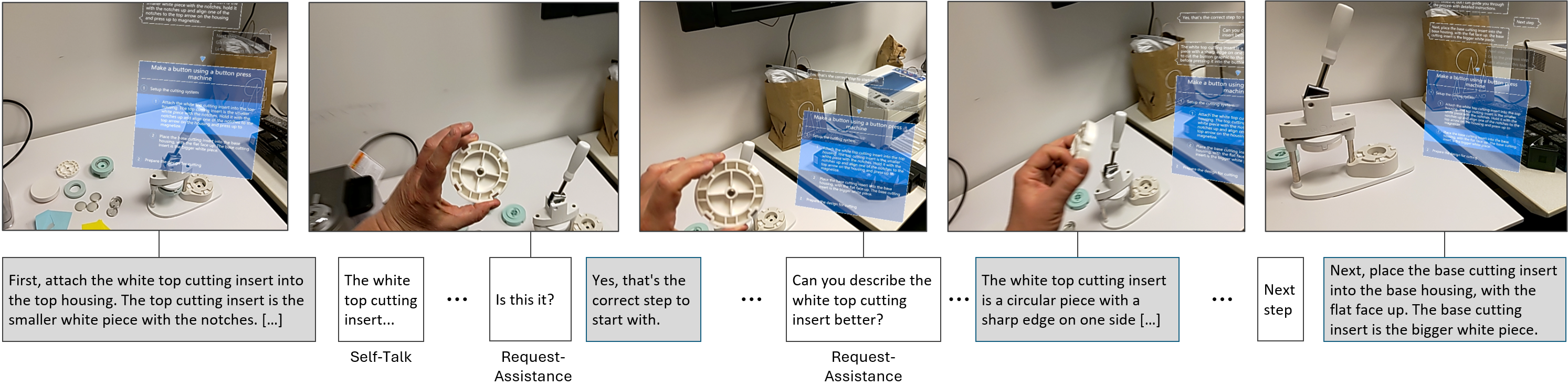}
  \caption{Segment from a participant interacting with \textsc{Sigma}. \textsc{Sigma} utterances have gray background; participant utterances have white background. The temporal axis is not to scale.}
  \Description{Segment from a participant interaction with Sigma.}
  \label{fig:SigmaInteraction}
\end{figure*}

Numerous visual question answering (VQA) datasets have been created to probe models' multimodal capabilities and to track their performance. Recent datasets extend beyond static image-based tasks and start to incorporate video, 3D, embodiment, and simulated environments for spatial and temporal reasoning. Some of the datasets \cite{mmbench,seed,sok,star} adopt post hoc question-answer (QA) generation from a third-person perspective as a way to probe the model's understanding of visual inputs. Other datasets \cite{openeqa,sq3d,leo} are generated in an embodied, situated environment to enable episodic and environment-specific reasoning. Several of these benchmark datasets were constructed by using LLMs \cite{seed,sok,leo} and crowd workers \cite{mmbench,sq3d}. In others, the authors themselves \cite{openeqa} generate hypothetical questions in simulated situations, along with the expected ground truth answers. Most of these datasets include only single-turn QAs, and they do not depict the realistic QAs that surface in a natural manner during real-time situated interaction, sometimes in the midst of long chains of multi-turn dialogues with a shared context and common ground that accrues over time.

% \begin{table}[h]
%   \centering
%   \small
%   \begin{tabular}{c|c|c|c|c}
%   \hline
%   \hline
%   \textbf{Dataset} & \textbf{Modality} & \textbf{Size} & \textbf{QA Gen} & \textbf{Q Type}\\
%   \hline\hline
%   MMBench\cite{mmbench} & Image & 3k & Human & MC \\
%   SEED-Bench\cite{seed}& Image+Video & 19k & LLM & MC \\
%   SOK-Bench\cite{sok} & Video & 44k & LLM+KG & MC \\
%   STAR\cite{star} & Video & 60k & Template+KG & MC\\
%   \hline
%   OpenEQA\cite{openeqa} & Video & 1.6k & Human & NL \\
%   SQA3D\cite{sq3d} & Video & 33.4k & Human & NL \\ 
%   LEO\cite{leo} & Image+3D & 1.5M & Human+LLM & NL\\\hline
%   SIGMA & Video & 113 & Interactive & NL\\
%   \hline
%     \hline
% \end{tabular}
% \caption{Visual Question Answering Datasets: {\normalfont Size = Number of Questions, MC = Multiple Choice, NL = Natural Language}}
% \label{vqadb}
% \end{table}
% \vspace{-10pt}

Several research projects have focused on the challenging task of using LLMs and LMMs to provide people with interactive guidance for procedural tasks. Some of these efforts \cite{argus,mrgs,dynamicmr,origami} focus on recognizing visual cues to help people to better understand task procedures. Other projects \cite{captaincook4d,errdec,industreal} emphasize recognizing steps, mistakes, objects, states, and gestures from egocentric videos. A few studies have proposed challenges to evaluate both visual and language capabilities for task assistance when considering synchronized multimodal streaming data in mixed reality \cite{holoassist, wtag}. The tasks considered in most benchmarks to date focus on the classification of dialog acts, in combination with high-level assessments of end-user satisfaction. To our knowledge, no studies have analyzed individual sentences to assess whether each question is answered correctly or addressed at the appropriate time and level of granularity. In distinction to prior research, we emphasize the importance of evaluating each question and answer in the context of the full situated collaborative history to that point, as captured during an interactive task guidance session with a real system. 

% \begin{table}[h]
%   \centering
%   \small
%   \begin{tabular}{l|c|c|c|l}
%   \hline\hline
%   \textbf{Dataset} & \textbf{Modality} & \textbf{Size} & \textbf{Instructor} & \textbf{Lan Eval}  \\
%   \hline\hline
%   Cook w/ Conv\cite{cookwconv} & Language & 48 & Human & Classification \\
%   Cook w/ Agent\cite{cookwagent} & Language & 16 & Human & Qualitative  \\
%   LLMR\cite{llmr} & XR & 11 & Model & Human Eval \\
%   WTaG\cite{wtag} & Multimodal & 56 & Human & Clsf+HmEvl  \\
%   HoloAssist\cite{holoassist} & Multimodal & 2221 & Human & Classification  \\ \hline
%   SIGMA & Multimodal & 26 & Model & QA Style  \\
%   \hline\hline
% \end{tabular}
% \caption{AI Assistant and Procedural Task Datasets: {\normalfont Size = Number of Recordings, Lan Eval = Language Evaluation Methods.}}
% \label{aitdb}
% \end{table}

\section{Interactive System-Driven Benchmarking}

Although constructing benchmarks by generating question-answer pairs post hoc on existing data can be useful for systematically probing the capabilities of LMMs with a battery of organized tests, we argue that the questions posed in such benchmarks lack ecological validity. That is, they do not reflect the types of questions end users pose in realistic scenarios, leading to a mismatch between measured model performance and the actual quality of user experiences.

EQA datasets can also be constructed from human-human interactions, where a participant takes on the role of an instructor or expert assistant (perhaps disguised as an AI system via Wizard-of-Oz paradigm) while another person carries out various tasks, asking questions along the way \cite{holoassist, wtag}
These questions capture the highest degree of ecological validity and might represent a gold standard for embodied question answering tasks. However, they are highly unstructured, contextual, and challenging to organize into a coherent benchmark.

Between the extremes of scenario-independent, artificial questions on one end and unconstrained human-to-human questions on the other, we propose an interactive system-driven approach to constructing ecologically valid situated assistance benchmarks. By building or leveraging an existing application, human-AI interactions can be collected that are situated in space and time within a task context.  As we will illustrate below, the challenges and questions that emerge from such interactions are distinct from those posed by existing benchmarks. These questions can be organized into a dataset, and models can be trained to more accurately answer questions from that dataset in the usual static benchmark paradigm. Importantly, these models can be further evaluated by integrating them back into the original application, and testing their abilities to improve user experience by answering questions in new, live interactions. These new interactions can surface new questions and challenges, within cycles of refinement.

\section{Data Collection}

We used \textsc{Sigma} \cite{bohus2024sigma, bohus2024sigmaarxiv}, an open-source mixed-reality task assistance system as the experimental test-bed for data collection. \textsc{Sigma} leverages a HoloLens 2 headset and uses speech recognition, speech synthesis, and LLMs to guide users step-by-step through procedural tasks in the physical world. Throughout the interaction, the system displays a virtual panel with instructions and reads the steps aloud. Users can navigate among steps and they can ask questions, which are in turn answered by prompting an LLM along with the current step context. A sample interaction snippet is illustrated in Figure \ref{fig:SigmaInteraction} and more system details are available in \cite{bohus2024sigma,bohus2024sigmaarxiv}.

Our long-term goal is to build a situated-assistance benchmark from data collected from users interacting with \textsc{Sigma} on a wide variety of tasks. We report here on a pilot data collection, with the short-term goals of refining the experimental protocol and identifying an initial set of salient challenges and phenomena. We plan to surface these challenges, along with others yet to be discovered, in the eventual benchmark.

For the pilot experiment, we authored and used five task recipes: resetting a Vertuo coffee machine to factory defaults, making coffee with a Nespresso coffee machine, changing the hard drive in a computer, creating craft buttons with a button making machine, and creating a notebook with a notebook binding machine. The experiment was reviewed and approved by our organization's IRB. After obtaining informed consent, each participant was briefed on how to use \textsc{Sigma}, carried out a short eye calibration sequence, and was presented with final instructions and safety warnings. Each participant carried out 2-4 tasks in a controlled lab space, then was debriefed and compensated \$50 for each hour of participation. Below, we report results from 26 interaction sessions collected, spanning the five defined tasks and nine participants.

\section{From EQA to Situated Assistance}

We present a qualitative analysis of 284 user utterances obtained from the data (after excluding easily parsable explicit navigation commands like \textit{``next step''}, and responses to when the system asks \textit{``Are you ready for the next step?''})

\subsection{Determining Response Obligation}

A first observation is that in situated-assistance settings, not all user utterances, even those phrased as questions, create an obligation for the system to respond. In fact, while the dataset contains many questions addressed to the system, a large proportion of utterances (42\%) reflect self-talk that users engage in as they perform steps (e.g., in Figure \ref{fig:SigmaInteraction}). While some self-talk utterances may provide an opportunity for the system to interject in an appropriate manner, they do not necessarily create an obligation to respond. At other times, users make statements that, while not in interrogative form, have the force of a question and release the conversational floor towards the system as a request for help, such as when reporting an unexpected error condition or an unexpected state, e.g., \textit{``I don't see a maximum fill line''} and \textit{``This is not going in.''} 

To better understand the space of user utterances, we manually annotated them with one of four possible dialog acts: \textbf{Request-Assistance} (38\%)---which creates an obligation for the system to respond; \textbf{Acknowledgment} (7\%) to a system utterance; \textbf{Self-Talk} (42\%); and \textbf{Step-Transition} (13\%)---in which the user reports that they are ready to transition to the next step or wish to hear a previous step again.

The data highlights that an important capability for situated assistive systems is knowing not just how to answer, but \emph{whether} a given user utterance should be answered in the first place. This task is not as simple as it may appear, and a first challenge can be formulated around it. In many instances, this problem cannot be resolved from text alone, but rather requires leveraging audio and visual information. As one example, there are 33 utterances for which the speech recognition result was \textit{``Ok''}. Of these, 12 denote the completion of the step, 11 are self-talk, and 10 are acknowledgments. Each usage requires a different response (or no response), and telling them apart requires careful consideration of the timing, prosody, and history of actions performed.

\subsection{Situated Question Answering}

An analysis of the \textbf{Request-Assistance} utterances in the collected data reveals that they differ in content and construction from the questions typically encountered in EQA challenge benchmarks. The differences stem largely from the fact that in an interactive setting, the participant and the system have shared attention, memory, and goals. They continuously interact to ground with each other and build mutual understanding over time. As a result, questions arise from real needs, are frequently specific, and are deeply enmeshed in the moment-by-moment flow of the collaboration. 

\subsubsection{Situated Questions}

We found that most of the user's questions aim to resolve grounding problems between the user and system regarding states, objects, and actions. A few other questions serve as general requests for help and conversation management. Several examples are shown in Table \ref{tab:utterances}. 

Often, questions center on resolving language-grounding issues, where the participant does not yet have an understanding of a particular term used by the system, e.g., \textit{``What is the base housing again?''}, or \textit{``What is the pin line?''}. In other cases, the participants understand the language, but need more details to fully ground the system's guidance to a concrete physical object, state or action, e.g., \textit{``How sharp is it supposed to be?''} or \textit{``How much should I fill?''}. Sometimes they simply elicit more information, e.g., \textit{``What does the capsule look like?''} or they aim to verify a specific grounding via a situated reference, e.g., \textit{``Is the white base the, the thing I just put on?''} Yet other times they aim to disambiguate between multiple possible groundings, e.g., \textit{``Is it the blue capsule or the brown capsule?''}.

\begin{table}
  \caption{Examples of Request-Assistance Utterances by Type}
  \label{tab:utterances}
  \begin{tabular*}{\columnwidth}{p{1.5cm}p{6.3cm}}
    \toprule
    \textbf{Type} & \textbf{Utterance} \\
    \midrule
    \small Grounding & \small ``Should the handle bar be up or down at this point?'' \\
    \small States & \small ``But the ca- capsule is coming out when I try to close it.''\\
    & \small ``Mm-hmm. But the blue capsule is too big to fit in.''\\
    \midrule
    \small Grounding & \small ``What is the base housing again?''\\
    \small Objects & \small ``Is this it?''\\
    & \small ``How sharp is it supposed to be?''\\
    & \small ``Is the white base the ... the thing I just put on?''\\
    \midrule
    \small Grounding & \small ``How much should I fill?''\\    
    \small Actions & \small ``Do I place, uh, the loop inside the holes?''\\
    & \small ``I understand rotating the ... base insert into place. What is rotating the top?''\\ 
    \midrule
    \small General Help & \small ``I need help, I'm lost.''\\
    \midrule
    \small Conv. Mgmt. & \small ``Can you say that last part again?''\\
    \bottomrule
\end{tabular*}
\end{table}

Many of the questions are anchored in the physical context and contain deictic expressions. Referring expressions for states, objects, and actions often contain personal and demonstrative pronouns such as \textit{\textbf{``it''}}, \textit{\textbf{``this''}}, or \textit{\textbf{``that''}} as in \textit{``How sharp is \textbf{it} supposed to be?''} or \textit{``\textbf{This} is not going in.''} More complex referring expressions also arise; e.g., in \textit{``Is the white base the ... \textbf{the thing I just put on}?''} where an object is referred to indirectly by indicating it as an object of a past action. The questions contain both \textit{endophoric} and \textit{exophoric} references; in the former the referent is available in previous utterances, whereas in the latter it is only available in the physical, multimodal context. Sometimes, these references are even combined. In one case, the participant held an object up as he asked the question \textit{``Is \textbf{this} \textbf{it}?''}. In this context, \textit{\textbf{``it''}} is an endophoric reference to the top cutting insert object that the system named in a previous utterance (\textit{``Attach \textbf{the top cutting insert} [...]''}) whereas \textit{\textbf{``this''}} is an exophoric reference to the object the user is actually holding in his hand. We emphasize the critical importance and rich set of research challenges captured by this single example. Situated assistive systems must handle such questions by reasoning about the full, multimodal context, including language and visual history over time. In turn, informative benchmarks should include these phenomena, as they naturally arise in the stream of interaction. 

Beyond text and vision, the prosodical contours and audio features of the questions can be informative, and sometimes essential for parsing meaning (they are typically missing in EQA benchmarks where the questions are usually given as text). As we have already discussed, not all requests for help are interrogatives. Furthermore, many questions produced while interacting contain natural disfluencies, hesitations, false-starts, restarts (see again Table \ref{tab:utterances}). These phenomena exacerbate challenges with segmenting and identifying requests for help and understanding how and when to best respond. Prosodical features sometimes dictate sentence semantics, as we have seen in the many facets of \textit{``Ok''} discussed in the previous subsection. Situated assistive systems need to understand these nuances in the acoustic channel to respond appropriately, and informative benchmarks need to capture these phenomena.

Overall, our initial data collection highlights that the questions a situated-assistive system needs to respond to are markedly different in both content and form from questions typically observed in existing EQA benchmarks. Consider the contrast to OpenEQA \cite{openeqa}, a recently proposed EQA benchmark in which questions were constructed by asking humans to watch video traces through homes and generate question-answer pairs while imagining themselves to be users of an assistive system. A cursory inspection of the questions in the two datasets reveals many important differences: for example, the \textsc{Sigma} data contains more deictic pronoun constructions, i.e., \textit{``it''} appears in 21\% of utterances vs 3\% in OpenEQA. The \textit{``How?''} questions in OpenEQA elicit general action plans, like \textit{``How can I clean the room?''} or \textit{``How can I dry my towel?''}, whereas more fine-grained and targeted questions eliciting missing information about actions, objects and states arise in \textsc{Sigma} interactions, e.g.,  \textit{``How much force should I use on the handle?''} or \textit{``Should I use the small or the big capsule?''} Object-related questions have a different structure in the two datasets: OpenEQA contains probing computer vision questions for recognizing objects and understanding spatial references, e.g., \textit{``What is on the left side of the bed?''}. In contrast, the \textsc{Sigma} dataset contains many questions aimed at grounding a language referent, e.g., \textit{``What is the base housing again?''}

\subsubsection{Situated Answers}

Throughout the data collection process, the \textsc{Sigma} system used a large language model prompt to identify questions and generate answers. Constructing ecologically valid benchmarks for situated assistance requires not only providing good coverage for the space of questions, but also understanding what makes for a good answer. For instance, we have noticed anecdotally that regardless of the correctness of answers from the LLM, the generations are often generic and too long-winded for the interactive setting.
Like questions, good answers should leverage the grounding between the participants, make use of references, and be efficient and to the point.

\subsection{Proactive Situated Assistance}

Seamless assistance goes well beyond answering questions. As frequently observed in human-human interactive datasets \cite{wtag, holoassist}, a good collaborative partner not only responds to requests for help, but continuously monitors the situation at hand and proactively intervenes to correct or prevent mistakes, provide clarifications and confirmations, or offer advice and encouragement as needed.

To our knowledge, there are no existing benchmarks that aim to assess capabilities for useful, proactive, system-initiative interventions. The development of such benchmarks raises significant challenges. Such an effort will require novel metrics and evaluation techniques, as both timing, content, form, and their alignment play an important role in the quality of a system-initiated intervention. 

The grounding issues that are communicated by participant questions are often visible prior to the participant actually asking the question. Various \emph{streaming} inference tasks for tracking these issues in real-time should be defined as waypoints towards enabling proactive intervention capabilities. Examples include detecting in-stream when a state, object, or action grounding issue arises, detecting when a participant starts or stops performing an action specified in natural language, detecting whether an action is being performed correctly, etc. In addition to understanding what happens in the physical world, understanding cognitive states such as user confusion, frustration, confidence, and focus also plays an important role in generating useful interventions. Human collaborative partners can often track this information in stream by leveraging non-verbal behaviors such as gaze and hand movement patterns, as well as information leaked through self-talk. We plan to develop benchmarks that assess whether and how well large multimodal models can perform on these types of streaming inference tasks, on the road towards enabling proactive interventions.

\section{Conclusion and Future Work}

In pursuit of constructing ecologically valid benchmarks for assessing the capabilities of LMMs for situated collaboration tasks, we propose and investigate an interactive system-driven approach. An initial analysis of data from a pilot study shows that questions arising through interaction differ in form and content from questions typically found in current EQA benchmarks. New challenges arise around determining when the system has an obligation to respond, and around developing tasks and metrics for benchmarking capabilities for in-stream, proactive interventions.

Based on the lessons learned, we plan to conduct a larger scale data collection with the \textsc{Sigma} system, and to define and publicly release a set of benchmarks and metrics around these challenges. In conjunction with the data, problem formulations and metrics, we will conduct experiments and release results from an initial assessment of existing LMM models on these challenges, coupled with a subsequent error analysis.

Future work includes investigating more natural collaborative settings. While \textsc{Sigma} and other datasets like HoloAssist \cite{holoassist} and WTaG \cite{wtag} provide a rich source of naturalistic questions, they are still collected in controlled lab settings. We expect that questions asked by people as they coordinate with AI systems and other people \emph{in the wild} will surface additional challenges with the real-time interpretation of language and activities.

% \begin{acks}

% \end{acks}

%\clearpage
\bibliographystyle{ACM-Reference-Format}
\bibliography{sample-base}

% \appendix


\end{document}